\documentclass[11pt,a4paper]{article}

\usepackage{amssymb}
\usepackage{graphicx}
\usepackage{bm}

\usepackage{mathrsfs}
\usepackage{cite}

\newcommand*\xbar[1]{%
  \hbox{%
    \vbox{%
      \hrule height 0.5pt 
      \kern0.5ex
      \hbox{%
        \kern-0.1em
        \ensuremath{#1}%
        \kern-0.1em
      }%
    }%
  }%
}

\begin{document}

\title{Three-loop verification of the equations relating running of the gauge couplings in ${\cal N}=1$ SQCD+SQED}

\author{
O.V.Haneychuk and K.V.Stepanyantz\\
\\
{\small{\em Moscow State University}}, {\small{\em  Faculty of Physics, Department  of Theoretical Physics}}\\
{\small{\em 119991, Moscow, Russia}}\\
}

\maketitle

\begin{abstract}
We verify a recently derived equations relating the renormalization group running of two gauge couplings in ${\cal N}=1$ SQCD+SQED by the explicit three-loop calculation. It is demonstrated that these equations are really valid in the HD+MSL scheme. In other words, if a theory is regularized by higher covariant derivatives and the renormalization is made by minimal subtractions of logarithms, the analogs of the strong and electromagnetic gauge couplings do not run independently. However, in the $\overline{\mbox{DR}}$ scheme the considered equations do not hold starting from the three-loop order, where the scheme dependence becomes essential. Therefore, they are valid only for a certain set of the renormalization prescriptions. We prove that all of them can be obtained from the HD+MSL scheme by finite renormalizations which satisfy a special constraint and illustrate how this works in the three-loop approximation.
\end{abstract}

\section{Introduction}
\hspace*{\parindent}

Even ${\cal N}=1$ supersymmetry essentially restricts possible ultraviolet divergences. For instance, according to the famous nonrenormalization theorem \cite{Grisaru:1979wc}, the superpotential does not receive divergent quantum corrections, so that the renormalization of masses and Yukawa couplings in supersymmetric theories is related to the renormalization of chiral matter superfields. In particular, for certain renormalization prescriptions the Yukawa $\beta$-function can be expressed in terms of their anomalous dimension. Surprisingly, the gauge $\beta$-function is also related to the anomalous dimension of the matter superfields by the NSVZ equation \cite{Novikov:1983uc,Jones:1983ip,Novikov:1985rd,Shifman:1986zi}. This equation can also be considered as a nonrenormalization theorem and holds for some special renormalization prescriptions, which are usually called ``NSVZ schemes''. Note that the most popular $\overline{\mbox{DR}}$ scheme (when a theory is regularized by dimensional reduction \cite{Siegel:1979wq} and divergences are removed by modified minimal subtraction \cite{Bardeen:1978yd}) is not NSVZ \cite{Jack:1996vg,Jack:1996cn,Jack:1998uj,Harlander:2006xq,Mihaila:2013wma}. In the case of using the dimensional reduction an NSVZ scheme may be constructed by a specially tuned finite renormalization in each order of the perturbation theory. However, no such a tuning is required in the case of using the Slavnov's higher covariant derivative regularization \cite{Slavnov:1971aw,Slavnov:1972sq,Slavnov:1977zf} formulated in terms of ${\cal N}=1$ superfields \cite{Krivoshchekov:1978xg,West:1985jx}. The use of this regularization is a main ingredient for constructing the perturbative proof of the NSVZ equation. This proof is based on the non-renormalization of the triple gauge-ghost vertices \cite{Stepanyantz:2016gtk}, factorization of loop integrals into integrals of double total derivatives in the momentum space \cite{Stepanyantz:2019ihw} (see also \cite{Smilga:2004zr,Stepanyantz:2011jy}), and the summation of singularities \cite{Stepanyantz:2019lfm,Stepanyantz:2020uke}. Using this proof it is easy to obtain \cite{Kataev:2013eta,Stepanyantz:2020uke} that a certain family of the NSVZ schemes is obtained in all orders if a theory is regularized by higher derivatives and divergences are removed by minimal subtractions of logarithms. This renormalization prescription is called the HD+MSL scheme \cite{Shakhmanov:2017wji,Stepanyantz:2017sqg}. Note that the higher covariant derivative regularization is not uniquely defined, because, e.g., the higher derivative regulator function can be chosen in different ways. Therefore, minimal subtraction of logarithms can supplement various versions of this regularization, so that the HD+MSL prescription is also not unique. However, all HD+MSL schemes are NSVZ in all orders of the perturbation theory.

Some interesting consequences of the NSVZ equations can be deduced for theories with multiple gauge couplings. It turns out that in some of them the anomalous dimensions of the matter superfields can be eliminated, and it is possible to derive equations relating the running of different gauge (and Yukawa) couplings. For instance, for the Minimal Supersymmatric Standard Model (MSSM) using the NSVZ equations \cite{Shifman:1996iy} and the equations describing the renormalization of the Yukawa couplings one can construct two all-loop renormalization group invariants (RGIs) \cite{Rystsov:2024soq}. A simpler example of ${\cal N}=1$ SQCD+SQED was considered in \cite{Kataev:2024amm}. In the massless limit this theory is described by the action

\begin{eqnarray}\label{SQCD+SQED_Superfield_Action}
&& S = \frac{1}{2g^2}\,\mbox{Re}\,\mbox{tr}\int d^4x\,d^2\theta\, W^a W_a + \frac{1}{4e^2}\,\mbox{Re}\int d^4x\, d^2\theta\, \bm{W}^a \bm{W}_a\nonumber\\
&&\qquad\qquad\qquad\qquad\quad + \sum\limits_{\mbox{\scriptsize a}=1}^{N_f}\, \frac{1}{4}\int d^4x\, d^4\theta\, \Big(\phi_{\mbox{\scriptsize a}}^+ e^{2V + 2q_{\mbox{\scriptsize a}}\bm{V}}\phi_{\mbox{\scriptsize a}}
+ \widetilde\phi_{\mbox{\scriptsize a}}^+ e^{-2V^T - 2q_{\mbox{\scriptsize a}}\bm{V}} \widetilde\phi_{\mbox{\scriptsize a}}\Big),\qquad
\end{eqnarray}

\noindent
which is written in the manifestly supersymmetric form with the help of ${\cal N}=1$ superfields. This action is invariant under the transformations of the gauge group $G\times U(1)$. The gauge superfields corresponding to the subgroups $G$ and $U(1)$ are denoted by $V$ and $\bm{V}$, respectively. The chiral matter superfields $\phi_{\mbox{\scriptsize a}}$ and $\widetilde\phi_{\mbox{\scriptsize a}}$ belong to the representations $R$ and $\xbar{R}$ of the non-Abelian group $G$, have the opposite $U(1)$ charges $q_{\mbox{\scriptsize a}} e$ and $-q_{\mbox{\scriptsize a}} e$, respectively, and form $N_f$ matter flavors. Here $e$ is the Abelian coupling constant, and the non-Abelian coupling constant is denoted by $g$. We will also use the notations $\alpha\equiv e^2/4\pi$ and $\alpha_s \equiv g^2/4\pi$. In the gauge part of the action $V = g V^A t^A$, where $t^A$ are the generators of the fundamental representation of the group $G$. In the matter part of the action $V = g V^A T^A$, where $T^A$ are the generators of the representation $R$. The generators satisfy the conditions

\begin{eqnarray}
&& \mbox{tr}(t^A t^B) = \frac{1}{2}\delta^{AB};\qquad\ \ \mbox{tr}(T^A T^B) = T(R)\delta^{AB};\quad\nonumber\\
&& [t^A, t^B]=if^{ABC} t^C;\qquad [T^A,T^B] = if^{ABC} T^C,
\end{eqnarray}

\noindent
where $f^{ABC}$ are the structure constants for the group $G$. For a simple group $G$ and an irreducible representation $R$

\begin{equation}
f^{ACD} f^{BCD} = C_2 \delta^{AB};\qquad (T^A T^B)_i{}^j = C(R)\cdot \delta_i^j.
\end{equation}

\noindent
In what follows the dimension of the representation $R$ will be denoted by $\mbox{dim}\,R$, so that $T(R) = C(R)\cdot\mbox{dim}\,R/r$, where $r \equiv \mbox{dim}\,G = \delta^{AA}$.

If $q_{\mbox{\scriptsize a}}=1$ for all $\mbox{a}$, then the anomalous dimensions of all matter superfields are the same. If, moreover, the representation $R$ is irreducible, then it is possible to eliminate the anomalous dimension of the matter superfields from the NSVZ equations and obtain the exact equation relating the renormalization group behaviour of the running gauge couplings \cite{Kataev:2024amm},

\begin{equation}\label{Beta_Relation}
\Big(1- \frac{\alpha_s C_2}{2\pi}\Big) \frac{\beta_s(\alpha_s,\alpha)}{\alpha_s^2} = -\frac{3C_2}{2\pi} + \frac{T(R)}{\mbox{dim}\,R}\cdot \frac{\beta(\alpha_s,\alpha)}{\alpha^2},
\end{equation}

\noindent
where the $\beta$-functions are defined by the equations

\begin{equation}\label{Beta_Definitions}
\beta(\alpha_s,\alpha) = \frac{d\alpha}{d\ln\mu}\bigg|_{\alpha_{s0},\alpha_0=\mbox{\scriptsize const}};\qquad \beta_s(\alpha_s,\alpha) = \frac{d\alpha_s}{d\ln\mu}\bigg|_{\alpha_{s0},\alpha_0=\mbox{\scriptsize const}},
\end{equation}

\noindent
and $\mu$ is a renormalization point. Note that these $\beta$-functions are scheme dependent starting from the three-loop approximation and do not depend on a renormalization prescription in the one- and two-loop approximations.

The relation (\ref{Beta_Relation}) can equivalently be rewritten in the form

\begin{equation}\label{RGI}
\Big(\frac{\alpha_s}{\mu^3}\Big)^{C_2} \exp\Big(\frac{2\pi}{\alpha_s} - \frac{T(R)}{\mbox{dim}\,R}\cdot \frac{2\pi}{\alpha}\Big) = \mbox{RGI},
\end{equation}

\noindent
where the acronym RGI in the right hand side means that this expression does not depend on a renormalization point $\mu$.

If the $U(1)$ charges $q_{\mbox{\scriptsize a}}$ are different for different flavors, then it is only possible to relate the $\beta$-function of ${\cal N}=1$ SQCD to the Adler $D$-function \cite{Adler:1974gd} defined by the equation\footnote{The detailed discussion of various definitions of the Adler function can be found in \cite{Aleshin:2019yqj}.}

\begin{equation}\label{SQCD+SQED_Adler_Definition}
D(\alpha_s)\equiv -\frac{3\pi}{2}\frac{d}{d\ln\mu}\Big(\frac{1}{\alpha}\Big)\bigg|_{\alpha_{s0},\alpha_{0}=\mbox{\scriptsize const},\ \alpha\to 0} = \frac{3\pi}{2} \lim\limits_{\alpha\to 0} \frac{\beta(\alpha_s,\alpha)}{\alpha^2},
\end{equation}

\noindent
which encodes quantum corrections to the Abelian gauge coupling due to the non-Abelian interaction. Using the exact NSVZ-like relation for this function derived in \cite{Shifman:2014cya,Shifman:2015doa} and the NSVZ equation
for ${\cal N}=1$ SQCD one can eliminate the anomalous dimension of the matter superfields and obtain the all-loop relation

\begin{equation}\label{Beta_Adler_Relation}
\beta_s(\alpha_s) = - \frac{\alpha_s^2}{2\pi(1-C_{2} \alpha_s/2\pi)} \bigg[\, 3 C_2 - \Big(\sum\limits_{\mbox{\scriptsize a}=1}^{N_f} q_{\mbox{\scriptsize a}}^2\Big)^{-1}\cdot\frac{4\, T(R) N_f D(\alpha_s)}{3\,\mbox{dim}\,R }  \bigg].
\end{equation}

For the MSSM the three-loop analysis (based on the calculations made in \cite{Jack:2004ch} and \cite{Haneychuk:2022qvu}) demonstrated that the renormalization group invariance of the expressions analogous to (\ref{RGI}) takes place for the HD+MSL renormalization prescription and is not valid in the $\overline{\mbox{DR}}$ scheme \cite{Rystsov:2024soq}. That is why we expect that Eqs. (\ref{Beta_Relation}) and (\ref{Beta_Adler_Relation}) are also satisfied only for some special renormalization prescriptions. In particular, they should be valid in the HD+MSL scheme. Really, these equations are derived from the NSVZ relations, while the HD+MSL scheme is NSVZ in all orders. In this paper we perform the explicit three-loop calculation of the renormalization group functions (RGFs) entering Eqs. (\ref{Beta_Relation}) and (\ref{Beta_Adler_Relation}) for various renormalization prescriptions supplementing the higher covariant derivative regularization. The arbitrariness of the choice of the renormalization scheme is encoded in a set of parameters fixing a way of removing divergences. The $\beta$-functions depend on these parameters  beyond the two-loop approximation. Using the three-loop results for them we find the constraints on a renormalization prescription under which Eqs. (\ref{Beta_Relation}) and (\ref{Beta_Adler_Relation}) are valid. We also present the expressions for relevant three-loop RGFs in the $\overline{\mbox{DR}}$ scheme and demonstrate that Eqs. (\ref{Beta_Relation}) and (\ref{Beta_Adler_Relation}) are not satisfied in this case.

The paper is organized as follows. In Sect. \ref{Section_Regularization} we briefly describe how the theory (\ref{SQCD+SQED_Superfield_Action}) is regularized by higher covariant derivatives. Next, in Sect. \ref{Section_Three_Loops_General} we verify Eq. (\ref{Beta_Relation}) which relates two gauge $\beta$-functions in the case of equal electromagnetic charges of all flavors. First, in Subsect. \ref{Subsection_General_HD+MSL} we present the three-loop expressions for both $\beta$-functions in the case of using a general renormalization prescription supplementing the higher covariant derivative regularization. Substituting these expressions into Eq. (\ref{Beta_Relation}) we find the values of finite constants fixing a subtraction scheme for which this equation holds. In particular, it is demonstrated that it is valid for the HD+MSL scheme(s). In Subsect. \ref{Subsection_General_DR} the three-loop $\beta$-functions are obtained in the $\overline{\mbox{DR}}$ scheme. We see that they satisfy Eq. (\ref{Beta_Relation}) only in the one- and two-loop approximations and do not satisfy it in the three-loop order. This implies that the $\overline{\mbox{DR}}$ scheme does not belong to the class of the renormalization prescriptions for which this equation holds. Various subtraction schemes of this class are related to each other (and, in particular, to the HD+MSL scheme) by finite renormalizations which satisfy a certain constraint constructed in Subsect. \ref{Subsection_General_Class}. The theory with different electromagnetic charges of different flavors is considered in Sect. \ref{Section_Three_Loops_Limit}. To verify the exact equation (\ref{Beta_Adler_Relation}), we compare the three-loop $\beta$-function of ${\cal N}=1$ SQCD with the Adler $D$-function for various renormalization prescriptions. As earlier, we see that Eq. (\ref{Beta_Adler_Relation}) is valid for the family of the HD+MSL schemes and does not hold in the $\overline{\mbox{DR}}$ scheme. The results are summarized in Conclusion.

\section{The higher derivative regularization of ${\cal N}=1$ SQCD+SQED}
\hspace*{\parindent}\label{Section_Regularization}

To regularize the theory (\ref{SQCD+SQED_Superfield_Action}) by higher covariant derivatives, first, we should add to the action a higher derivative term. It is also expedient to use the background field method \cite{DeWitt:1965jb,Arefeva:1974jv,Abbott:1980hw,Abbott:1981ke}, which ensures manifest ${\cal N}=1$ supersymmetry of the effective action. In the superfield formulation \cite{Gates:1983nr,West:1990tg,Buchbinder:1998qv} it can be introduced with the help of the replacement $e^{2V} \to e^{2{\cal F}(v)} e^{2V}$, where $V$ in the right hand side becomes the (non-Abelian) background gauge superfield, and $v$ is the quantum gauge superfield. The function ${\cal F}(v)$ is needed for taking into account the nonlinear renormalization of this quantum gauge superfield \cite{Piguet:1981fb,Piguet:1981hh,Tyutin:1983rg}.\footnote{The explicit form of this function in the lowest nontrivial order has been found in \cite{Juer:1982fb,Juer:1982mp}, where its necessity for making multiloop calculations was explicitly demonstrated. Moreover, it was demonstrated that without this function the renormalization group equations are not satisfied \cite{Kazantsev:2018kjx}.} It is also necessary to replace the couplings by the bare ones, which in our notations are marked by the subscript 0. After that, the regularized action takes the form

\begin{eqnarray}\label{SQCD+SQED_Regularized_Action}
&& S_{\mbox{\scriptsize reg}} = \frac{1}{2g_0^2}\,\mbox{Re}\,\mbox{tr}\int d^4x\,d^2\theta\, W^a \Big[e^{-2V} e^{-2{\cal F}(v)} R\Big(-\frac{\xbar{\nabla}^2\nabla^2}{16\Lambda^2}\Big)\, e^{2{\cal F}(v)} e^{2V}\Big]_{\mbox{\scriptsize Adj}} W_a\nonumber\\
&& + \frac{1}{4e_0^2}\,\mbox{Re}\int d^4x\, d^2\theta\, \bm{W}^a R\Big(\frac{\partial^2}{\Lambda^2}\Big) \bm{W}_a
+ \sum\limits_{\mbox{\scriptsize a}=1}^{N_f}\, \frac{1}{4}\int d^4x\, d^4\theta\, \Big(\phi_{\mbox{\scriptsize a}}^+ e^{2{\cal F}(v)} e^{2V + 2q_{\mbox{\scriptsize a}}\bm{V}}\phi_{\mbox{\scriptsize a}}\qquad
\nonumber\\
&& + \widetilde\phi_{\mbox{\scriptsize a}}^+ e^{-2V^T - 2q_{\mbox{\scriptsize a}}\bm{V}} e^{-2{\cal F}(v)^T}  \widetilde\phi_{\mbox{\scriptsize a}}\Big).\qquad\vphantom{\frac{1}{2}}
\end{eqnarray}

\noindent
Here the sum of the usual supersymmetric Yang--Mills (SYM) action and the higher derivative term generates the function $R$ such that $R(0)=1$ and $R(x)\to \infty$ at $x\to\infty$. Note that, following \cite{Korneev:2021zdz}, we use the same regulator function for the non-Abelian and Abelian parts of the SYM action. The right and left covariant derivatives present in the action (\ref{SQCD+SQED_Regularized_Action}) (for the conventions adopted in this paper) are defined as

\begin{equation}
\nabla_a = D_a;\qquad \xbar\nabla_{\dot a} = e^{2{\cal F}(v)} e^{2V} \xbar D_{\dot a} e^{-2V} e^{-2{\cal F}(v)},
\end{equation}

\noindent
respectively. The subscript $\mbox{Adj}$ means that in the expressions marked by it one should use the generators of the adjoint representation of the gauge group, so that

\begin{equation}
\big(f_0 + f_1 V + f_2 V^2 +\ldots\big)_{\mbox{\scriptsize Adj}} X \equiv f_0 X + f_1 [V,X] + f_2 [V,[V,X]] + \ldots
\end{equation}

\noindent
In the kinetic term for the Abelian gauge superfield the function $R$ depends on the usual derivatives, because the corresponding superfield strength $\bm{W}_a$ is gauge invariant. The theory under consideration does not contain Yukawa couplings, and, therefore, there is no need to introduce the higher covariant derivatives in the matter part of the action. However, the higher covariant derivatives are also introduced in the gauge fixing terms.

The higher derivative term removes divergences beyond the one-loop approximation. However, according to \cite{Slavnov:1977zf}, for regularizing the remaining one-loop divergences it is necessary to insert the Pauli--Villars determinants into the generating functional. For supersymmetric gauge theories with a simple gauge group the set of the Pauli--Villars determinants has been constructed in \cite{Aleshin:2016yvj,Kazantsev:2017fdc}. The generalization to the case of multiple gauge couplings has been done in \cite{Korneev:2021zdz}. In particular, to regularize the theory (\ref{SQCD+SQED_Superfield_Action}), for the non-Abelian subgroup $G$ (of the gauge group $G\times U(1)$) we need to introduce three chiral Pauli--Villars superfields $\varphi_{1,2,3}$ in the adjoint representation and the chiral superfields $\Phi_G$ and $\widetilde\Phi_G$ in the representations $R$ and $\xbar R$. All these superfields have the $U(1)$ charges equal to 0. The superfields $\varphi_{1,2,3}$ with the mass $M_\varphi$ cancel one-loop divergences (and subdivergences) coming from diagrams with a loop of the quantum gauge superfield or ghosts. The superfields $\Phi_G$ and $\widetilde\Phi_G$ with the mass $M_G$ cancel the divergent one-loop contributions to the renormalization of the non-Abelian coupling constant coming from a matter loop. The actions for the Pauli--Villars superfields corresponding to the subgroup $G$ have the form

\begin{eqnarray}
&&\hspace*{-3mm} S_\varphi = \frac{1}{2} \mbox{tr} \int d^4x\,d^4\theta \Big(\varphi_1^+ \Big[\, R\Big(-\frac{\xbar{\nabla}^2\nabla^2}{16\Lambda^2}\Big)\, e^{2{\cal F}(v)} e^{2V}\Big]_{\mbox{\scriptsize Adj}} \varphi_1 + \varphi_2^+ (e^{2{\cal F}(v)} e^{2V})_{\mbox{\scriptsize Adj}} \varphi_2\nonumber\\
&&\hspace*{-3mm}\qquad\qquad\qquad\qquad + \varphi_3^+ (e^{2{\cal F}(v)} e^{2V})_{\mbox{\scriptsize Adj}} \varphi_3 \Big) + \bigg[\frac{1}{2}\mbox{tr} \int d^4x\,d^2\theta\, M_\varphi \Big(\varphi_1^2 + \varphi_2^2 + \varphi_3^2\Big) + \mbox{c.c.}\bigg];\quad\nonumber\\
&&\hspace*{-3mm} S_{\Phi_G} = \frac{1}{4} \int d^4x\,d^4\theta \Big(\Phi_G^+\, e^{2{\cal F}(v)} e^{2V} \Phi_G + \widetilde\Phi_G^+\, e^{-2{\cal F}(v)^T} e^{-2V^T} \widetilde\Phi_G\Big)\nonumber\\
&&\hspace*{-3mm}\qquad\qquad\qquad\qquad\qquad\qquad\qquad\qquad\qquad\ + \bigg[\frac{1}{2} \int d^4x\,d^2\theta\, M_G\, \widetilde\Phi_G^T\, \Phi_G +\mbox{c.c.}\bigg].\quad
\end{eqnarray}

\noindent
For the $U(1)$ subgroup of the gauge group it is sufficient to use the chiral Pauli--Villars superfields $\Phi_1$ and $\widetilde\Phi_1$ in the trivial representation of the subgroup $G$ with the $U(1)$ charges $+e$ and $-e$, respectively. They have the mass $M_1$ and are described by the action

\begin{equation}
S_{\Phi_1} = \frac{1}{4} \int d^4x\,d^4\theta \Big(\Phi_1^+ e^{2\bm{V}} \Phi_1 + \widetilde\Phi_1^+ e^{-2\bm{V}} \widetilde\Phi_1\Big) + \bigg[\frac{1}{2} \int d^4x\,d^2\theta\, M_1 \widetilde\Phi_1^T \Phi_1 +\mbox{c.c.}\bigg].
\end{equation}

It is essential that all Pauli--Villars masses should be proportional to the cutoff $\Lambda$ present in the higher derivative term,

\begin{equation}
M_\varphi = a_\varphi \Lambda;\qquad M_G = a_G \Lambda;\qquad M_1 = a_1\Lambda,
\end{equation}

\noindent
where the parameters $a_\varphi$, $a_G$, and $a_1$ do not depend on couplings. Their values can be chosen in an arbitrary way. As we will see in what follows, the three-loop $\beta$-functions will depend on these parameters together with the parameter

\begin{equation}
A \equiv \int\limits_0^\infty dx\,\ln x\,\frac{d}{dx}\frac{1}{R(x)}.
\end{equation}

\noindent
The value of this parameter is determined by the higher derivative regulator function $R(x)$ present in Eq. (\ref{SQCD+SQED_Regularized_Action}). For instance, for the simplest regulator $R(x) = 1+x^n$, where $n$ is a positive integer, this parameter vanishes, $A=0$.

The generating functional $Z$ is defined by the equation

\begin{eqnarray}\label{Generating_Z}
&& Z[\mbox{Sources}] = \int D\mu\, \mbox{Det}(PV, M_G)\, \mbox{Det}^{\sum q_{\mbox{\tiny a}}^2}(PV,M_1)\nonumber\\
&&\qquad\qquad\qquad\qquad \times \exp\Big(iS_{\mbox{\scriptsize reg}} + i S_{\mbox{\scriptsize gf}} + i S_{\mbox{\scriptsize FP}} + i S_{\mbox{\scriptsize NK}} + i S_\varphi + i S_{\mbox{\scriptsize Sources}}\Big),
\end{eqnarray}

\noindent
where $D\mu$ denotes the measure of the functional integration, $S_{\mbox{\scriptsize gf}}$ is the gauge fixing term, $S_{\mbox{\scriptsize FP}}$ and $S_{\mbox{\scriptsize NK}}$ are the actions for the Faddev--Popov and Nielsen--Kallosh ghosts, respectively, and $S_{\mbox{\scriptsize sources}}$ includes all relevant sources. The Pauli--Villars determinants entering the expression (\ref{Generating_Z}) are defined as

\begin{equation}
\mbox{Det}^{-1}(PV, M_G) \equiv \int D\Phi_G D\widetilde\Phi_G \exp(i S_{\Phi_G});\quad \mbox{Det}^{-1}(PV, M_1) \equiv \int D\Phi_1 D\widetilde\Phi_1 \exp(i S_{\Phi_1}).
\end{equation}

\noindent
The powers of these determinants in Eq. (\ref{Generating_Z}) are chosen in such a way that the loops of the Pauli--Villars superfields do cancel the one-loop divergences.

\section{Three-loop verification of the equation relating two $\beta$-functions}
\hspace*{\parindent}\label{Section_Three_Loops_General}

In this section we consider the theory (\ref{SQCD+SQED_Superfield_Action}) in the particular case $q_{\mbox{\scriptsize a}}=1$ for all $\mbox{a}=1,\ldots, N_f$ and verify the exact expression (\ref{Beta_Relation}) by the explicit three-loop calculation. In Subsect. \ref{Subsection_General_HD+MSL} we present the three-loop expressions for both $\beta$-functions in the case of using the higher covariant derivative regularization supplemented by an arbitrary renormalization prescription, and demonstrate that Eq. (\ref{Beta_Relation}) is valid if the finite constants fixing a subtraction scheme satisfy certain constraints. In the HD+MSL scheme these constraints are satisfied as they should be. In Subsect. \ref{Subsection_General_DR} we present the results for the $\beta$-functions in the $\overline{\mbox{DR}}$ scheme and demonstrate that Eq. (\ref{Beta_Relation}) does not hold for this renormalization prescription starting from the three-loop approximation. In Subsect. \ref{Subsection_General_Class} we show that any two renormalization schemes for that Eq. (\ref{Beta_Relation}) is satisfied are related by a finite renormalization which obeys a certain constraint.

\subsection{General renormalization prescription supplementing the HD regularization}
\hspace*{\parindent}\label{Subsection_General_HD+MSL}

For the theory (\ref{SQCD+SQED_Superfield_Action}) with $q_{\mbox{\scriptsize a}}=1$ for all $\mbox{a}=1,\ldots, N_f$ we calculate the three-loop $\beta$-functions with the help of the general expression constructed in \cite{Haneychuk:2025ejd} for theories with multiple gauge couplings regularized by higher covariant derivatives.\footnote{A similar expression for theories with a single gauge coupling has been obtained in \cite{Kazantsev:2020kfl} on the base of the general statements proved in \cite{Stepanyantz:2020uke}. Although at present not all parts of this expression have been checked directly, it exactly agrees with the results of all three-loop calculations done so far with the higher covariant derivative regularization \cite{Shakhmanov:2017soc,Kazantsev:2018nbl,Kuzmichev:2019ywn,Aleshin:2020gec,Aleshin:2022zln}.} The one- and two-loop contributions to the gauge $\beta$-functions are scheme independent and can be calculated immediately. However, starting from the three-loop approximation the $\beta$-functions depend on a way of removing divergences in the previous orders or, in other words, on a renormalization prescription. In each order of the perturbation theory this arbitrariness leads to the emergence of some finite constants. Fixing values of these constants one sets a renormalization prescription. That is why before presenting the three-loop $\beta$-functions, it is necessary to discuss a subtraction scheme used in the calculation.

First, for the theory under consideration we obtain the (scheme independent) two-loop $\beta$-functions and, after that, integrate the renormalization group equations (\ref{Beta_Definitions}). The result can be presented in the form of the relations between the bare and renormalized coupling constants,

\begin{eqnarray}\label{Renormalization_Prescription_General_Alpha}
&&\hspace*{-5mm} \frac{1}{\alpha_0} = \frac{1}{\alpha} - \frac{N_f\,\mbox{dim}\,R}{\pi} \Big(\ln\frac{\Lambda}{\mu} + d_1\Big)
- \frac{\alpha_s}{\pi^2} N_f C(R)\, \mbox{dim}\,R \Big(\ln\frac{\Lambda}{\mu} + d_{2}\Big) - \frac{\alpha}{\pi^2} N_f \,\mbox{dim}\, R \nonumber\\
&&\hspace*{-5mm} \times \Big(\ln\frac{\Lambda}{\mu} + \widetilde d_2\Big)  + O(\alpha_s^2,\alpha_s\alpha,\alpha^2);\\
&&\hspace*{-5mm} \vphantom{1}\nonumber\\
\label{Renormalization_Prescription_General_AlphaS}
&&\hspace*{-5mm} \frac{1}{\alpha_{s0}} = \frac{1}{\alpha_s} + \frac{3C_2}{2\pi}\Big(\ln\frac{\Lambda}{\mu} + b_{11}\Big) - \frac{N_f T(R)}{\pi} \Big(\ln\frac{\Lambda}{\mu} + b_{12}\Big)
+ \frac{3\alpha_s}{4\pi^2} (C_2)^2 \Big(\ln\frac{\Lambda}{\mu} + b_{21}\Big)\nonumber\\
&&\hspace*{-5mm}  - \frac{\alpha_s}{2\pi^2} N_f C_2 T(R) \Big(\ln\frac{\Lambda}{\mu} + b_{22}\Big) - \frac{\alpha_s}{\pi^2} N_f C(R) T(R) \Big(\ln\frac{\Lambda}{\mu} + b_{23}\Big) - \frac{\alpha}{\pi^2} N_f T(R) \Big(\ln\frac{\Lambda}{\mu}\nonumber\\
&&\hspace*{-5mm} + \widetilde b_{21}\Big) + O(\alpha_s^2,\alpha_s\alpha,\alpha^2),\vphantom{\frac{1}{2}}
\end{eqnarray}

\noindent
where the bare couplings are marked by the subscript $0$. The integration constants $b_i$, $d_i$, etc. determine a renormalization prescription in the lowest orders. By definition, in the HD+MSL scheme the renormalization constants include only the terms containing $\ln^k\Lambda/\mu$ with $k\ge 1$. This in particular implies that the couplings satisfy the boundary conditions

\begin{equation}
\alpha^{-1}\Big|_{\mu=\Lambda} = (Z_\alpha \alpha_0)^{-1}\Big|_{\mu=\Lambda} = \alpha_0^{-1};\qquad \alpha_s^{-1}\Big|_{\mu=\Lambda} = (Z_{\alpha_s} \alpha_{s0})^{-1}\Big|_{\mu=\Lambda} = \alpha_{s0}^{-1},
\end{equation}

\noindent
analogous to the ones discussed in \cite{Kataev:2013eta}. Therefore, in the considered approximation the HD+MSL scheme is obtained if $b_{11}=b_{12}=b_{21}=b_{22}=b_{23}=\widetilde b_{23}=\widetilde b_{21}=0$, $d_1=d_2=\widetilde d_2=0$.

The $\beta$-functions depend on the parameters fixing the renormalization prescription starting from the three-loop approximation, where the scheme dependence manifests itself. Namely, the three-loop results for the gauge $\beta$-functions of the model (\ref{SQCD+SQED_Superfield_Action}) with $q_{\mbox{\scriptsize a}}=1$ for all $\mbox{a}=1,\ldots, N_f$ calculated with the help of the general formula presented in \cite{Haneychuk:2025ejd} have the form

\begin{eqnarray}\label{Beta_Alpha}
&&\hspace*{-5mm} \frac{\beta(\alpha_s,\alpha)}{\alpha^2} = \frac{N_f\, \mbox{dim}\, R}{\pi}\bigg\{1+ \frac{\alpha}{\pi}
+ \frac{\alpha_s}{\pi} C(R) - \frac{1}{2\pi^2}\Big(\alpha+\alpha_s C(R)\Big)^2
- \frac{\alpha^2}{\pi^2} N_f\,\mbox{dim}\,R \Big(\ln a_1\nonumber\\
&&\hspace*{-5mm} + 1 + \frac{A}{2} + \widetilde d_2 - d_1\Big)
+ \frac{3\alpha_s^2}{2\pi^2} C_2 C(R) \Big(\ln a_\varphi + 1 + \frac{A}{2} + d_2 - b_{11}\Big)
- \frac{\alpha_s^2}{\pi^2} N_f C(R) T(R) \nonumber\\
&&\hspace*{-5mm} \times \Big(\ln a_G + 1 +\frac{A}{2} + d_{2} - b_{12}\Big) + O(\alpha_s^3,\alpha_s^2\alpha,\alpha_s\alpha^2,\alpha^3)\bigg\};\\
&&\hspace*{-5mm} \vphantom{1}\nonumber\\
\label{Beta_Alpha_S}
&&\hspace*{-5mm} \frac{\beta_s(\alpha_s,\alpha)}{\alpha_s^2} = - \frac{1}{2\pi}\Big(3 C_2 - 2 N_f T(R)\Big) + \frac{\alpha}{\pi^2} N_f T(R) + \frac{\alpha_s}{4\pi^2}\Big(-3(C_2)^2 + 2 N_f C_2 T(R)\nonumber\\
&&\hspace*{-5mm} + 4 N_f C(R) T(R)\Big) - \frac{\alpha^2}{\pi^3} (N_f)^2 T(R)\, \mbox{dim}\, R \Big(\ln a_1 + 1 + \frac{A}{2} + \widetilde b_{21} - d_1\Big) - \frac{1}{2\pi^3} N_f T(R)\nonumber\\
&&\hspace*{-5mm} \times \Big(\alpha+\alpha_s C(R)\Big)^2 + \frac{\alpha\alpha_s}{2\pi^3} N_f C_2 T(R) - \frac{3\alpha_s^2}{8\pi^3} (C_2)^3 \Big(1 + 3b_{21} - 3b_{11}\Big) + \frac{\alpha_s^2}{4\pi^3} N_f (C_2)^2
\nonumber\\
&&\hspace*{-5mm} \times T(R) \Big(1 + 3b_{21} - 3b_{11} + 3 b_{22} - 3b_{12}\Big) + \frac{3\alpha_s^2}{2\pi^3} N_f C_2 C(R) T(R) \Big(\ln a_\varphi + \frac{4}{3} + \frac{A}{2}\nonumber\\
&&\hspace*{-5mm} + b_{23} - b_{11}\Big) - \frac{\alpha_s^2}{2\pi^3} (N_f)^2 C_2 T(R)^2 (b_{22} - b_{12})
- \frac{\alpha_s^2}{\pi^3} (N_f)^2 C(R) T(R)^2 \Big(\ln a_G + 1\nonumber\\
&&\hspace*{-5mm} + \frac{A}{2} + b_{23} - b_{12}\Big) + O(\alpha_s^3,\alpha_s^2\alpha,\alpha_s\alpha^2,\alpha^3).
\end{eqnarray}

\noindent
These expressions are written for the case of using the higher covariant derivative regularization supplemented by the renormalization prescription given by Eqs. (\ref{Renormalization_Prescription_General_Alpha}) and (\ref{Renormalization_Prescription_General_AlphaS}).

Substituting the expressions (\ref{Beta_Alpha}) and (\ref{Beta_Alpha_S}) into Eq. (\ref{Beta_Relation}) we see that it is valid if the finite constants fixing a subtraction scheme obey the constraints

\begin{equation}\label{Finite_Constants_Constraints_General}
\widetilde b_{21} = \widetilde d_2;\qquad b_{21} = b_{11};\qquad b_{22} = b_{12};\qquad b_{23} = d_2.
\end{equation}

\noindent
Evidently, if all these constants vanish, then these relations are satisfied. In other words, the equations (\ref{Finite_Constants_Constraints_General}) and, therefore, the relation (\ref{Beta_Relation}) in the three-loop approximation are valid for the HD+MSL renormalization prescription for arbitrary values of the regularization parameters $a_\varphi$, $a_G$, $a_1$, and $A$. Note that this check is nontrivial because in the three-loop approximation the scheme dependence becomes essential.

The explicit dependence of Eqs. (\ref{Beta_Alpha}) and (\ref{Beta_Alpha_S}) on the renormalization parameters allows constructing a subtraction scheme in which the relation (\ref{Beta_Relation}) is valid and RGFs are as simple as possible. For this purpose we choose such values of finite constants that the constraints (\ref{Finite_Constants_Constraints_General}) are satisfied and, moreover, impose the conditions

\begin{equation}
b_{12} = b_{11} + \ln \frac{a_G}{a_\varphi};\qquad
d_2 = b_{11} - \ln a_\varphi - 1 - \frac{A}{2};\qquad
\widetilde d_2 = d_1 - \ln a_1 - 1 -\frac{A}{2}.
\end{equation}

\noindent
In this case the expressions for the $\beta$-functions take the simplest form

\begin{eqnarray}\label{Beta_Alpha_Minimal}
&&\hspace*{-7mm} \frac{\beta(\alpha_s,\alpha)}{\alpha^2} = \frac{N_f\, \mbox{dim}\, R}{\pi}\bigg\{1+ \frac{\alpha}{\pi}
+ \frac{\alpha_s}{\pi} C(R) - \frac{1}{2\pi^2}\Big(\alpha+\alpha_s C(R)\Big)^2 + O(\alpha_s^3,\alpha_s^2\alpha,\alpha_s\alpha^2,\alpha^3)\bigg\};\nonumber\\
&&\hspace*{-5mm} \vphantom{1}\\
\label{Beta_Alpha_S_Minimal}
&&\hspace*{-7mm} \frac{\beta_s(\alpha_s,\alpha)}{\alpha_s^2} = - \frac{1}{2\pi}\Big(3 C_2 - 2 N_f T(R)\Big) + \Big(1+\frac{\alpha_s C_2}{2\pi}\Big)\bigg[\frac{\alpha}{\pi^2} N_f T(R) + \frac{\alpha_s}{4\pi^2}\Big(-3(C_2)^2
+ 2 N_f \nonumber\\
&&\hspace*{-5mm}  \times C_2 T(R) + 4 N_f C(R) T(R)\Big)\bigg] - \frac{1}{2\pi^3} N_f T(R) \Big(\alpha+\alpha_s C(R)\Big)^2  + O(\alpha_s^3,\alpha_s^2\alpha,\alpha_s\alpha^2,\alpha^3).
\end{eqnarray}

\noindent
The resulting scheme is analogous to the ``minimal'' scheme constructed in \cite{Shirokov:2022jyd} for ${\cal N}=1$ SQED with $N_f$ flavors. In particular, we see that in the $\beta$-function for the Abelian coupling constant all terms proportional to $(N_f)^k$ with $k\ge 2$ disappear exactly as for ${\cal N}=1$ SQED.

\subsection{$\overline{\mbox{DR}}$ scheme}
\hspace*{\parindent}\label{Subsection_General_DR}

Although the higher covariant derivative regularization has a lot of attractive features, especially in the supersymmetric case \cite{Stepanyantz:2019lyo}, most calculations for such theories have been done with the help of the dimensional reduction \cite{Siegel:1979wq} supplemented by modified minimal subtraction \cite{Bardeen:1978yd}.\footnote{The dimensional regularization \cite{'tHooft:1972fi,Bollini:1972ui,Ashmore:1972uj,Cicuta:1972jf} explicitly breaks supersymmetry \cite{Delbourgo:1974az}.} That is why it is also desirable to present the $\beta$-functions for this renormalization prescription, which is usually called the $\overline{\mbox{DR}}$ scheme. In fact, the result in the $\overline{\mbox{DR}}$ scheme can be derived from Eqs. (\ref{Beta_Alpha}) and (\ref{Beta_Alpha_S}) if one takes into account that the arbitrariness in the choice of a renormalization prescription can always compensate the arbitrariness in the choice of a regularization. This implies that it is possible to reproduce the $\overline{\mbox{DR}}$ expressions by a proper choice of the finite constants in Eqs. (\ref{Renormalization_Prescription_General_Alpha}) and (\ref{Renormalization_Prescription_General_AlphaS}). The corresponding values of these constants have been found in \cite{Kazantsev:2020kfl} for a theory with a single gauge coupling and in \cite{Haneychuk:2022qvu} for the case of multiple gauge couplings. For the theory under consideration these values are

\begin{eqnarray}\label{Finite_Constants_DR_General}
&&\hspace*{-5mm} d_1 = \ln a_1;\qquad b_{11} = \ln a_\varphi;\qquad b_{12} = \ln a_G;\qquad \widetilde d_2 = -\frac{1}{4}-\frac{A}{2};\qquad d_2 = -\frac{1}{4}-\frac{A}{2};\nonumber\\
&&\hspace*{-5mm} \widetilde b_{21} = -\frac{1}{4} - \frac{A}{2};\qquad b_{21} = \frac{1}{4} + \ln a_\varphi;\qquad b_{22} = \frac{1}{4} + \ln a_G;\qquad b_{23} = -\frac{1}{4} - \frac{A}{2}.
\end{eqnarray}

\noindent
Substituting them into Eqs. (\ref{Beta_Alpha}) and (\ref{Beta_Alpha_S}) we obtain the results for the $\beta$-functions in the $\overline{\mbox{DR}}$ scheme,

\begin{eqnarray}\label{Beta_Alpha_DR}
&& \frac{\beta(\alpha_s,\alpha)}{\alpha^2}\bigg|_{\mbox{\scriptsize $\overline{\mbox{DR}}$}} = \frac{N_f\, \mbox{dim}\, R}{\pi}\bigg\{1+ \frac{\alpha}{\pi}
+ \frac{\alpha_s}{\pi} C(R) - \frac{1}{2\pi^2}\Big(\alpha+\alpha_s C(R)\Big)^2
- \frac{3\alpha^2}{4\pi^2} N_f\,\mbox{dim}\,R \nonumber\\
&& + \frac{9\alpha_s^2}{8\pi^2} C_2 C(R) - \frac{3\alpha_s^2}{4\pi^2} N_f C(R) T(R) + O(\alpha_s^3,\alpha_s^2\alpha,\alpha_s\alpha^2,\alpha^3)\bigg\};\\
&& \vphantom{1}\nonumber\\
\label{Beta_Alpha_S_DR}
&& \frac{\beta_s(\alpha_s,\alpha)}{\alpha_s^2}\bigg|_{\mbox{\scriptsize $\overline{\mbox{DR}}$}} = - \frac{1}{2\pi}\Big(3 C_2 - 2 N_f T(R)\Big) + \frac{\alpha}{\pi^2} N_f T(R) + \frac{\alpha_s}{4\pi^2}\Big(-3(C_2)^2 + 2 N_f C_2 T(R)\nonumber\\
&& + 4 N_f C(R) T(R)\Big) - \frac{3\alpha^2}{4\pi^3} (N_f)^2 T(R)\,\mbox{dim}\, R - \frac{1}{2\pi^3} N_f T(R) \Big(\alpha+\alpha_s C(R)\Big)^2 + \frac{\alpha\alpha_s}{2\pi^3} N_f C_2 \nonumber\\
&& \times T(R) - \frac{21\alpha_s^2}{32\pi^3} (C_2)^3 + \frac{5\alpha_s^2}{8\pi^3} N_f (C_2)^2 T(R) + \frac{13\alpha_s^2}{8\pi^3} N_f C_2 C(R) T(R) - \frac{\alpha_s^2}{8\pi^3} (N_f)^2 C_2 T(R)^2\nonumber\\
&& - \frac{3\alpha_s^2}{4\pi^3} (N_f)^2 C(R) T(R)^2 + O(\alpha_s^3,\alpha_s^2\alpha,\alpha_s\alpha^2,\alpha^3).
\end{eqnarray}

\noindent
In particular, we see that the dependence on the higher derivative regularization parameters vanished as it should. Certainly, this can be considered as a test of the calculation correctness.

Substituting the expressions (\ref{Beta_Alpha_DR}) and (\ref{Beta_Alpha_S_DR}) into Eq. (\ref{Beta_Relation}) we see that the relation (\ref{Beta_Relation}) is not satisfied starting from the three-loop approximation,

\begin{eqnarray}
&& \Big(1- \frac{\alpha_s C_2}{2\pi}\Big) \frac{\beta_s(\alpha_s,\alpha)}{\alpha_s^2} +\frac{3C_2}{2\pi} - \frac{T(R)}{\mbox{dim}\,R}\cdot \frac{\beta(\alpha_s,\alpha)}{\alpha^2}\nonumber\\
&&\qquad\qquad\qquad = -\frac{9\alpha_s^2}{32\pi^3} (C_2)^3 + \frac{3\alpha_s^2}{8\pi^3} (C_2)^2 N_f T(R) + O(\alpha_s^3,\alpha_s^2\alpha,\alpha_s\alpha^2,\alpha^3). \qquad
\end{eqnarray}

\noindent
This is not unexpected because it is well known that the $\overline{\mbox{DR}}$ scheme is not NSVZ \cite{Jack:1996vg,Jack:1996cn,Jack:1998uj,Harlander:2006xq,Mihaila:2013wma}, while Eq. (\ref{Beta_Relation}) has been derived from the NSVZ equations. The fact that Eq. (\ref{Beta_Relation}) does not hold in the $\overline{\mbox{DR}}$ scheme can also be derived by comparing the values of finite constants (\ref{Finite_Constants_DR_General}) corresponding to this renormalization prescription with the equations (\ref{Finite_Constants_Constraints_General}), some of them in this case are not satisfied.

\subsection{Renormalization prescriptions for which the equation (\ref{Beta_Relation}) is valid}
\hspace*{\parindent}\label{Subsection_General_Class}

Earlier we saw that Eq. (\ref{Beta_Relation}) is satisfied only for certain renormalization prescriptions which include the HD+MSL scheme(s) and do not include the $\overline{\mbox{DR}}$ scheme. Therefore, there is a class of renormalization prescriptions for that Eq. (\ref{Beta_Relation}) holds. This class can be described similarly to the set of the NSVZ schemes \cite{Goriachuk:2018cac,Goriachuk_Conference,Goriachuk:2020wyn}. Namely, we start with an HD+MSL scheme and construct all other renormalization prescriptions with the help of finite redefinitions of couplings. For the schemes in that Eq. (\ref{Beta_Relation}) is valid these finite renormalizations should satisfy a certain constraint. Really, let us assume that there are two such renormalization prescriptions. The gauge coupling constants for the first one will be denoted by $\alpha$ and $\alpha_s$. For the second subtraction scheme the gauge couplings will be denoted by $\alpha'$ and $\alpha_s'$. According to \cite{Vladimirov:1975mx,Vladimirov:1979my} these two sets of gauge couplings can be related by a finite renormalization

\begin{equation}\label{Finite_Renormalization_General}
\alpha' = \alpha'(\alpha_s,\alpha);\qquad \alpha_s' = \alpha_s'(\alpha_s,\alpha).
\end{equation}

Let us consider the first subtraction scheme. Substituting the expressions (\ref{Beta_Definitions}) into Eq. (\ref{Beta_Relation}) and integrating the resulting equation we obtain the relation \cite{Kataev:2024amm}

\begin{equation}\label{First_Scheme}
\frac{1}{\alpha_s} -\frac{1}{\alpha_{s0}} + \frac{C_2}{2\pi} \ln \frac{\alpha_s}{\alpha_{s0}} - \frac{T(R)}{\mbox{dim}\,R} \Big(\frac{1}{\alpha} - \frac{1}{\alpha_0}\Big) = C - \frac{3C_2}{2\pi}\ln\frac{\Lambda}{\mu},
\end{equation}

\noindent
where $C$ is a constant. The analogous relation should also be valid for the second renormalization prescription,

\begin{equation}\label{Second_Scheme}
\frac{1}{\alpha_s'} -\frac{1}{\alpha_{s0}} + \frac{C_2}{2\pi} \ln \frac{\alpha_s'}{\alpha_{s0}} - \frac{T(R)}{\mbox{dim}\,R} \Big(\frac{1}{\alpha'} - \frac{1}{\alpha_0}\Big) = C'- \frac{3C_2}{2\pi}\ln\frac{\Lambda}{\mu},
\end{equation}

\noindent
where $C'$ is (in general) another constant. Subtracting Eq. (\ref{First_Scheme}) from Eq. (\ref{Second_Scheme}) we see that the finite renormalization (\ref{Finite_Renormalization_General}) should satisfy the constraint

\begin{equation}\label{Restriction_General}
\frac{1}{\alpha_s'} - \frac{1}{\alpha_s} + \frac{C_2}{2\pi}\ln \frac{\alpha_s'}{\alpha_s} = \frac{T(R)}{\mbox{dim}\,R}\Big(\frac{1}{\alpha'} - \frac{1}{\alpha}\Big) + B,
\end{equation}

\noindent
where $B\equiv C'-C$ is a constant.

Thus, any two subtraction schemes in which Eq. (\ref{Beta_Relation}) holds are related by a finite renormalization (\ref{Finite_Renormalization_General}) that satisfies the constraint (\ref{Restriction_General}). In particular, any such scheme is related to the HD+MSL scheme by such a finite renormalization.

To illustrate this, we construct a finite renormalization relating the HD+MSL scheme (which in the lowest orders is obtained if $d_1=d_2=\widetilde d_2 = b_{11} = b_{12} = b_{21}=b_{22} = b_{23}=\widetilde b_{21}=0$) to the scheme given by Eqs. (\ref{Renormalization_Prescription_General_Alpha}) and (\ref{Renormalization_Prescription_General_AlphaS}). The couplings in the latter scheme will now be denoted by primes, while the couplings in the HD+MSL scheme will be denoted by $\alpha$ and $\alpha_s$. Then in the three-loop approximation the finite renormalization we are interested in can be written in the form

\begin{eqnarray}\label{Finite_Renormalization_Three_Loops_General_Alpha}
&&\hspace*{-11mm} \frac{1}{\alpha'} = \frac{1}{\alpha} + \frac{N_f\,\mbox{dim}\,R}{\pi} d_{1} + \frac{\alpha_s}{\pi^2} N_f C(R)\, \mbox{dim}\,R\, d_2 + \frac{\alpha}{\pi^2} N_f \,\mbox{dim}\, R\, \widetilde d_2 + O(\alpha_s^2,\alpha_s\alpha,\alpha^2);\\
&&\hspace*{-11mm} \vphantom{1}\nonumber\\
\label{Finite_Renormalization_Three_Loops_General_AlphaS}
&&\hspace*{-11mm} \frac{1}{\alpha_{s}'} = \frac{1}{\alpha_s} - \frac{3C_2}{2\pi} b_{11} + \frac{N_f T(R)}{\pi} b_{12}  - \frac{3\alpha_s}{4\pi^2} (C_2)^2 b_{21} + \frac{\alpha_s}{2\pi^2} N_f C_2 T(R) b_{22}
\nonumber\\
&&\hspace*{-11mm}\qquad\qquad\qquad\qquad\qquad\quad + \frac{\alpha_s}{\pi^2} N_f C(R) T(R) b_{23} + \frac{\alpha}{\pi^2} N_f T(R) \widetilde b_{21}
+ O(\alpha_s^2,\alpha_s\alpha,\alpha^2).
\end{eqnarray}

\noindent
Substituting these expressions into Eq. (\ref{Restriction_General}) and equating the coefficients at the same powers of couplings and at the same group factors we see that the finite constants in Eqs. (\ref{Finite_Renormalization_Three_Loops_General_Alpha}) and (\ref{Finite_Renormalization_Three_Loops_General_AlphaS}) should satisfy the constraints

\begin{eqnarray}\label{Relations_Between_Finite_Constants_General}
&& B = - \frac{3C_2}{2\pi} b_{11} + \frac{N_f T(R)}{\pi} b_{12} - \frac{N_f T(R)}{\pi} d_{1};\nonumber\\
&& b_{21} = b_{11};\qquad b_{22} = b_{12};\qquad b_{23} = d_2;\qquad \widetilde b_{21} = \widetilde d_2.
\end{eqnarray}

\noindent
The first equation here determines the value of $B$, while the other equations coincide with Eq. (\ref{Finite_Constants_Constraints_General}) obtained earlier by a different method. This coincidence can be considered as a correctness test for Eqs. (\ref{Finite_Constants_Constraints_General}) and (\ref{Restriction_General}).

\section{Three-loop verification of the relation between the ${\cal N}=1$ SQCD $\beta$-function and the Adler $D$-function}
\hspace*{\parindent}\label{Section_Three_Loops_Limit}

In this section we consider the theory (\ref{SQCD+SQED_Superfield_Action}) assuming that the Abelian charges $q_{\mbox{\scriptsize a}}$ may be different for different values of $\mbox{a}$. In the limit $\alpha\to 0$ its $\beta$-function $\beta_s(\alpha_s,\alpha\to 0)$ coincides with the $\beta$-function of ${\cal N}=1$ SQCD and can be related to the Adler $D$-function by Eq. (\ref{Beta_Adler_Relation}). Here we verify this relation in the three-loop approximation for various renormalization prescriptions supplementing the higher covariant derivative regularization and in the $\overline{\mbox{DR}}$ scheme.

The three-loop $\beta$-function for ${\cal N}=1$ SQCD regularized by higher covariant derivatives can be obtained either from the result of Ref. \cite{Kazantsev:2020kfl}, or from Eq. (\ref{Beta_Alpha_S}) after taking the limit $\alpha\to 0$. The result is written as

\begin{eqnarray}\label{General_Beta}
&&\hspace*{-5mm} \beta_s(\alpha_s) = -\frac{\alpha_s^2}{2\pi}\Big(3 C_2 - 2 N_f T(R)\Big) + \frac{\alpha_s^3}{4\pi^2}\Big(-3 (C_2)^2 + 2 N_f C_2 T(R) + 4 N_f C(R)\, T(R)\Big)
\nonumber\\
&&\hspace*{-5mm} + \frac{\alpha_s^4}{8\pi^3} \bigg[-3 (C_2)^3 \Big(1+3b_{21}-3b_{11}\Big) + 2 N_f (C_2)^2\, T(R)\Big(1+3b_{21}-3b_{12}+3b_{22}-3b_{11}\Big)\nonumber\\
&&\hspace*{-5mm} - 4 N_f C(R)^2\, T(R) - 4 (N_f)^2 C_2 T(R)^2 \Big(b_{22} - b_{12}\Big) + 4 N_f C_2\, C(R)\,T(R)\Big(3\ln a_\varphi + 4 + \frac{3A}{2} \nonumber\\
&&\hspace*{-5mm} + 3b_{23} - 3b_{11}\Big) - 8 (N_f)^2 C(R) T(R)^2 \Big(\ln a_G + 1 + \frac{A}{2} + b_{23} - b_{12}\Big)\bigg] + O(\alpha_s^5)
\end{eqnarray}

\noindent
and depends on the regularization parameters $a_G$, $a_\varphi$, $A$\footnote{Note that for pure ${\cal N}=1$ SQCD there is no need to introduce the Pauli--Villars superfields $\Phi_1$ and $\widetilde\Phi_1$, so that the parameter $a_1$ is absent in this case.} and the finite constants $b_i$ which determine the renormalization prescription. In the case of ${\cal N}=1$ SQCD (without the Abelian gauge superfield) these constants are defined by Eq. (\ref{Renormalization_Prescription_General_AlphaS}) in which it is necessary to take the limit $\alpha\to 0$.

The three-loop Adler $D$-function for an arbitrary renormalization prescription supplementing the higher covariant derivative regularization has been calculated in \cite{Kataev:2017qvk}. In the notations adopted in this paper the result takes the form

\begin{eqnarray}\label{General_Adler}
&&\hspace*{-5mm} D(\alpha_s) = \frac{3}{2} \sum\limits_{\mbox{\scriptsize a}=1}^{N_f} q_{\mbox{\scriptsize a}}^2\, \mbox{dim}\, R\, \bigg\{1 + \frac{\alpha_s}{\pi} C(R) -\frac{\alpha_s^2}{2\pi^2} C(R)^2 + \frac{3\alpha_s^2}{2\pi^2} C_2 C(R) \Big(\ln a_\varphi+1+\frac{A}{2}
\nonumber\\
&&\hspace*{-5mm} +d_2-b_{11}\Big) - \frac{\alpha_s^2 N_f}{\pi^2} C(R)\, T(R)\,\Big(\ln a_G + 1 + \frac{A}{2} + d_2 - b_{12}\Big) + O(\alpha_s^3)\bigg\}
\end{eqnarray}

\noindent
and also depends on a new finite constant $d_2$ defined by the renormalization of the electromagnetic coupling,

\begin{equation}\label{Alpha_Renormalization}
\lim\limits_{\alpha\to 0}\Big(\frac{1}{\alpha} - \frac{1}{\alpha_0}\Big) = \frac{1}{\pi} \sum\limits_{\mbox{\scriptsize a}=1}^{N_f} q_{\mbox{\scriptsize a}}^2\,\mbox{dim}\, R \bigg[\Big(\ln\frac{\Lambda}{\mu} + d_1\Big) + \frac{\alpha_s}{\pi} C(R) \Big(\ln\frac{\Lambda}{\mu} + d_2\Big) \bigg] + O(\alpha_s^2).
\end{equation}

\noindent
This equation is similar to Eq. (\ref{Renormalization_Prescription_General_Alpha}), but the right hand side depends on $q_{\mbox{\scriptsize a}}$ and does not contain the coupling $\alpha$.

Comparing Eqs. (\ref{General_Beta}) and (\ref{General_Adler}) we conclude that in the considered approximation the equation (\ref{Beta_Adler_Relation}) holds if the finite constants which determine the renormalization prescription obey the conditions

\begin{equation}\label{Finite_Constants_Constraints}
b_{21} = b_{11};\qquad b_{22} = b_{12};\qquad b_{23} = d_2.
\end{equation}

In the HD+MSL scheme all finite constants are equal to 0, and these constraints are satisfied. Therefore, the relation (\ref{Beta_Adler_Relation}) is really valid in any HD+MSL scheme, as expected, since the underlying NSVZ equations are satisfied in this case in all orders for arbitrary values of regularization parameters.

The arbitrariness in the choice of parameters fixing a renormalization prescription can be used for constructing a scheme in which RGFs are simplified as much as possible provided that the exact relation (\ref{Beta_Adler_Relation}) holds in the considered approximation. The corresponding ``minimal'' scheme is obtained if the renormalization parameters satisfy Eq. (\ref{Finite_Constants_Constraints}) and the additional constraints

\begin{equation}
b_{12} = b_{11} + \ln \frac{a_G}{a_\varphi};\qquad d_2 = b_{11} - \ln a_\varphi - 1 - \frac{A}{2}.
\end{equation}

\noindent
In this case RGFs take the simplest possible form

\begin{eqnarray}\label{General_Beta_Minimal}
&&\hspace*{-8mm} \beta_s(\alpha_s) = -\frac{\alpha_s^2}{2\pi}\Big(3 C_2 - 2 N_f T(R)\Big) + \frac{\alpha_s^3}{4\pi^2}\Big(1+\frac{\alpha_s C_2}{2\pi}\Big)\Big(-3 (C_2)^2 + 2 N_f C_2 T(R)
\nonumber\\
&&\hspace*{-8mm}\qquad\qquad\qquad\qquad\qquad\qquad\qquad\ \ + 4 N_f C(R)\, T(R)\Big) - \frac{\alpha_s^4}{2\pi^3} N_f C(R)^2\, T(R) + O(\alpha_s^5);\\
\label{General_Adler_Minimal}
&&\hspace*{-8mm} D(\alpha_s) = \frac{3}{2} \sum\limits_{\mbox{\scriptsize a}=1}^{N_f} q_{\mbox{\scriptsize a}}^2\, \mbox{dim}\, R\, \bigg\{1 + \frac{\alpha_s}{\pi} C(R) -\frac{\alpha_s^2}{2\pi^2} C(R)^2 + O(\alpha_s^3)\bigg\}.
\end{eqnarray}

\noindent
(Note that in this paper we always assume that the renormalization prescriptions are compatible with  the structure of quantum corrections, so that no further simplification can be done.)

According to \cite{Kataev:2017qvk} and \cite{Aleshin:2019yqj}, the $\overline{\mbox{DR}}$ results for the $\beta$-function and for the Adler $D$-function are reproduced from Eq. (\ref{General_Beta}) and (\ref{General_Adler}) if the finite constants take the values

\begin{eqnarray}
&& b_{11} = \ln a_\varphi;\qquad\qquad\, b_{12} = d_1 = \ln a_G;\qquad\, b_{21} = \ln a_\varphi + \frac{1}{4};\qquad\nonumber\\
&& b_{22} = \ln a_G + \frac{1}{4};\qquad\, b_{23} = -\frac{1}{4} - \frac{A}{2};\qquad\quad d_2 = -\frac{1}{4}-\frac{A}{2}.
\end{eqnarray}

\noindent
In this case only the last constraint in Eq. (\ref{Finite_Constants_Constraints}) is satisfied, so that in the $\overline{\mbox{DR}}$ scheme the relation (\ref{Beta_Adler_Relation}) is not valid.

Repeating the reasoning of Subsect. \ref{Subsection_General_Class} it is easy to see that any two renormalization prescriptions for which Eq. (\ref{Beta_Adler_Relation}) is valid in all orders are related by the finite renormalization

\begin{equation}\label{Finite_Renormalization_Limit}
\alpha_s'=\alpha_s'(\alpha_s);\qquad (\alpha')^{-1} = \alpha^{-1} + f(\alpha_s)
\end{equation}

\noindent
which satisfies the constraint

\begin{equation}\label{Restriction_Adler}
\frac{1}{\alpha_s'(\alpha_s)} - \frac{1}{\alpha_s} + \frac{C_2}{2\pi}\ln \frac{\alpha_s'(\alpha_s)}{\alpha_s} = \Big(\sum\limits_{\mbox{\scriptsize a}=1}^{N_f} q_{\mbox{\scriptsize a}}^2\Big)^{-1}\cdot \frac{N_f T(R)}{\mbox{dim}\,R}\, f(\alpha_s) + B,
\end{equation}

\noindent
where $\alpha_s'(\alpha_s)$ and $f(\alpha_s)$ are finite functions of the non-Abelian coupling, and $B$ is a constant. Exactly as in Subsect. \ref{Subsection_General_Class}, it is possible to demonstrate that in the three-loop approximation this equation gives the constraints (\ref{Finite_Constants_Constraints}) together with the same expression for the constant $B$ as in Eq. (\ref{Relations_Between_Finite_Constants_General}).\footnote{However, it is necessary to remember that if the charges $q_{\mbox{\scriptsize a}}$ are different, then the finite constants $d_1$ and $d_2$ are defined by Eq. (\ref{Alpha_Renormalization}) (which slightly differs from analogous Eq. (\ref{Renormalization_Prescription_General_Alpha})).}

\section{Conclusion}
\hspace*{\parindent}

According to \cite{Kataev:2024amm}, the renormalization group running of two gauge couplings in ${\cal N}=1$ SQCD+SQED is related by the exact relations (\ref{Beta_Relation}) and (\ref{Beta_Adler_Relation}). In this paper we have made the verification of these relations by an explicit three-loop calculation. Note that in the one- and two-loop approximations the $\beta$-functions do not depend on a renormalization prescription. The scheme dependence becomes essential only in the three-loop approximation, so that this is a minimal order of the perturbation theory in which its effects can be tested. It was demonstrated that in the three-loop approximation Eqs. (\ref{Beta_Relation}) and (\ref{Beta_Adler_Relation}) are really valid in the HD+MSL scheme, when a theory is regularized by the Slavnov's higher covariant derivative method and the renormalization is performed with the help of minimal subtractions of logarithms. This result agrees with the general statement that the HD+MSL prescription gives an NSVZ scheme in all orders of the perturbation theory \cite{Stepanyantz:2020uke}, because Eqs. (\ref{Beta_Relation}) and (\ref{Beta_Adler_Relation}) were derived from the NSVZ relations. In the $\overline{\mbox{DR}}$ scheme the relations between RGFs of ${\cal N}=1$ SQCD+SQED  do not hold starting from the three-loop order, again, in agreement with the fact that the $\overline{\mbox{DR}}$ scheme is not NSVZ for theories with unextended supersymmetry \cite{Jack:1996vg,Jack:1996cn,Jack:1998uj}.

We also describe all renormalization prescriptions for that Eqs. (\ref{Beta_Relation}) and (\ref{Beta_Adler_Relation}) are valid in all orders. They are related to the HD+MSL scheme by finite renormalizations which satisfy the constraints (\ref{Restriction_General}) and (\ref{Restriction_Adler}). In the three-loop approximation these constraints lead to some relations between finite constants fixing a renormalization procedure, see Eq. (\ref{Relations_Between_Finite_Constants_General}) and (\ref{Finite_Constants_Constraints}). Requiring the validity of these constraints and using the remaining arbitrariness in choosing the parameters which fix a subtraction scheme (in the considered approximation) it is possible to construct such a renormalization prescription that the exact equations are satisfied and RGFs have the simplest form. The use of such ``minimal'' renormalization schemes may essentially simplifies various multiloop calculations with the higher derivative regularization, because, as usual, the most complicated integrals correspond to the scheme dependent constributions to RGFs, see, e.g. \cite{Shirokov:2022jyd}.

Thus, by an explicit three-loop calculation we have confirmed that in ${\cal N}=1$ SQCD+SQED for certain renormalization prescriptions (which include the HD+MSL and ``minimal'' schemes) the renormalization group running of the non-Abelian gauge coupling is really related to the renormalization group running of the Abelian one.

\section*{Acknowledgments}

The authors are very grateful to A.L.Kataev for valuable discussions.

\end{document}